\documentclass[aps,prl,twocolumn,superscriptaddress,floatfix]{revtex4-1}

\usepackage{graphicx}
\usepackage{amsmath}
\usepackage{bm}
\usepackage{color,soul}

\begin{document}
\setlength{\textfloatsep}{12pt}
%\preprint{}

\title{Harnessing Dispersion in Soliton Microcombs to Mitigate Thermal Noise}


\author{Jordan R. Stone}
\affiliation{Time and Frequency Division, National Institute for Standards and Technology, Boulder, CO 80305}
\affiliation{Department of Physics, University of Colorado Boulder, Boulder, CO 80309}
\email{jordan.stone@colorado.edu}

\author{Scott B. Papp}
\affiliation{Time and Frequency Division, National Institute for Standards and Technology, Boulder, CO 80305}
\affiliation{Department of Physics, University of Colorado Boulder, Boulder, CO 80309}

%\collaboration{Xu Yi}

\date{\today}

\begin{abstract}
We explore intrinsic thermal noise in soliton microcombs, revealing thermodynamic correlations induced by nonlinearity and group-velocity dispersion. A suitable dispersion design gives rise to control over thermal-noise transduction from the environment to a soliton microcomb. We present simulations with the Lugiato-Lefever equation (LLE), including temperature as a stochastic variable. By systematically tuning the dispersion, we suppress repetition-rate frequency fluctuations by up to $50$ decibels for different LLE soliton solutions. In an experiment, we observe a measurement-system-limited $15$-decibel reduction in the repetition-rate phase noise for various settings of the pump-laser frequency, and our measurements agree with a thermal-noise model. Finally, we compare two octave-spanning soliton microcombs with similar optical spectra and offset frequencies, but with designed differences in dispersion. Remarkably, their thermal-noise-limited carrier-envelope-offset frequency linewidths are 1 MHz and 100 Hz, which demonstrates an unprecedented potential to mitigate thermal noise. Our results guide future soliton-microcomb design for low-noise applications, and, more generally, they illuminate emergent properties of nonlinear, multi-mode optical systems subject to intrinsic fluctuations. 
\end{abstract}

%\pacs{}
%\keywords{}

\maketitle

%\section{}
\setlength{\parskip}{0em}
Optical-frequency combs are powerful and versatile tools for making precision measurements across the electromagnetic spectrum \cite{fortier201920}. 
To reach applications outside the laboratory, integrated-photonics frequency-combs based on continuous-wave (CW) laser-pumped microresonator solitons are rapidly being advanced \cite{kippenberg2018dissipative}. Microresonators simultaneously achieve high quality factor ($Q$) and small mode volume ($V$) to intensify the intraresonator field, enhance nonlinearity, and promote interactions between all the comb modes. On the one hand, a large $Q/V$ ratio enables experiments to access exotic nonlinear regimes \cite{cole2017soliton, anderson2017coexistence} and realize octave-spanning combs for applications, including clocks \cite{newman2018photonic, drake2018kerr} and optical-frequency synthesizers \cite{briles2018interlocking}. On the other hand, small $V$ increases the sensitivity to environmental and pump-laser fluctuations, which in turn degrades the comb coherence and application performance \cite{drake2019thermal,liu2020photonic}. Recently, a high-signal-to-noise measurement of the carrier-envelope-offset frequency ($f_{\rm{ceo}}$) revealed the thermal noise limit for soliton microcombs \cite{drake2019thermal}. Indeed, frequency fluctuations in the soliton-microcomb repetition rate, $f_{\rm{rep}}$, were observed to behave according to the fundamental thermodynamic relation \cite{landau1980statistical} 
\begin{equation}{\label{eq:FD}}
\left< \delta f_{\rm{rep}}^2 \right>=\lvert \eta_T \rvert^2\frac{k_BT^2}{\rho C V},
\end{equation}
where $\left< \delta f_{\rm{rep}}^2 \right>$ is the variance of $f_{\rm{rep}}$ frequency fluctuations, $\eta_T=\frac{df_{\rm{rep}}}{dT}$, $k_B$ is Boltzmann's constant, $T$ is the microresonator modal temperature, $\rho$ is the material density,  and $C$ is the specific heat. Hence, understanding $\eta_T$ is crucial for interpreting thermal noise and how soliton microcombs interact with their environment. 

\begin{figure}[h!]
    \centering
    \includegraphics[width=219 pt]{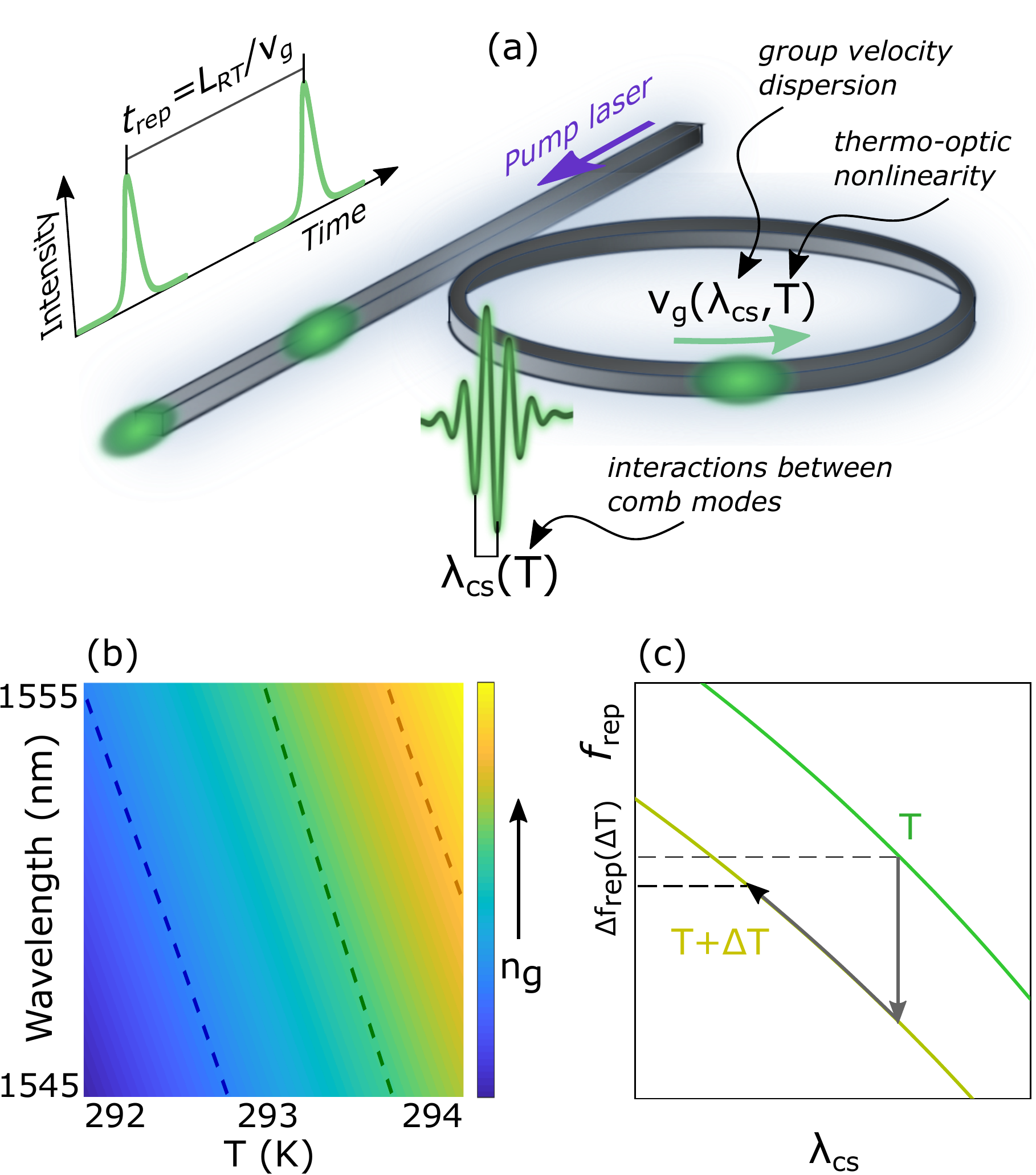}
    \caption{(a) The soliton velocity, $v_{\rm{g}}$, depends on both its wavelength, $\lambda_{\rm{cs}}$, due to group-velocity dispersion (GVD), and the modal temperature, $T$, due to the thermo-optic effect. Hence, the pulse-to-pulse timing, $t_{\rm{rep}}=1/f_{\rm{rep}}$, is sensitive to $T$. (b) GVD curves with $T$-dependence. Dashed lines are lines of constant $n_{\rm{g}}$. (c) Trajectory of $f_{\rm{rep}}$ for a change in $T$, $\Delta T$. The GVD curve shifts due to the thermo-optic effect (vertical arrow), and the correlated change in $\lambda_{\rm{CS}}$ moves $f_{\rm{rep}}$ along its GVD curve (arrow parallel to curve). The change in $f_{\rm{rep}}$, $\Delta f_{\rm{rep}}$, is the total vertical displacement.}
    \label{fig:Concept}
\end{figure}

Here, we present a comprehensive set of predictions and measurements on controlling thermal noise in soliton microcombs, and we reveal unique behaviors of thermal-noise correlations mapped to nonlinear light propagation. Since the soliton comb modes are phase-locked, they collectively respond to extrinsic properties dictated by the resonator, such as group-velocity dispersion (GVD or dispersion) and temperature. Hence, transduction of thermal fluctuations to $f_{\rm{rep}}$ fluctuations through $\eta_T$ is coupled to how the soliton is influenced by dispersion, which is readily controlled in integrated-photonics resonators. We present an experimental validation of our predictions, which is universally applicable to soliton microcombs. With a single resonator device, we observe that the thermal-noise-limited $f_{\rm{rep}}$ phase noise varies significantly with the pump-laser frequency, $\nu_{\rm{p}}$. Finally, we investigate the impact of these physics on $f_{\rm{ceo}}$; our simulations indicate that spectrally similar, octave-spanning soliton microcombs may feature significant differences in their thermal-noise-limited linewidth.   

%\hl{We show how interactions between the numerous comb modes dictate a collective response to changes in temperature that underlies the transduction of thermal noise to .} 
%In particular, we highlight the role of dispersion in determining this response. 

Temperature induces strong correlations in soliton microcombs (Fig. \ref{fig:Concept}), particularly the transduction of thermal noise to $f_{\rm{rep}}$ that is quantified by $\eta_T$. Figure \ref{fig:Concept}a shows how the thermo-optic effect connects $f_{\rm{rep}}$ to the microresonator temperature, $T$, according to 
\begin{equation}
f_{\rm{rep}}(\lambda_{\rm{CS}}, T)=\frac{v_{\rm{g}}(\lambda_{\rm{CS}}, T)}{L_{RT}}=\frac{\rm{c}}{n_{\rm{g}}(\lambda_{\rm{CS}}, T)L_{RT}},
\end{equation}
where $\lambda_{\rm{CS}}$ is the wavelength of the soliton carrier wave, $v_{\rm{g}}$ is the soliton group velocity, $L_{RT}$ is the microresonator circumference, $c$ is the speed of light in vacuum, and $n_{\rm{g}}$ is the group refractive index that depends on both $\lambda_{\rm{CS}}$ due to group-velocity dispersion (GVD) and $T$ due to the thermo-optic effect (Fig. \ref{fig:Concept}b). (In our analysis, we do not consider thermomechanical effects that couple $L_{RT}$ to $T$. While this approximation is justified \cite{matsko2007whispering}, thermomechanical effects can be included in our model in a straightforward way and would not impact our conclusions). Importantly, we discover that optical nonlinearity and GVD couple $\lambda_{\rm{CS}}$ to $T$, as depicted in Fig. \ref{fig:Concept}a, and we note that such effects are unique to nonlinear, multi-mode optical systems. Hence, temperature combines with $\lambda_{\rm{CS}}$ to determine the value of $\eta_T$. Figure \ref{fig:Concept}c offers a graphical interpretation of $\eta_T$ and depicts two GVD curves at temperatures $T$ and $T+\Delta T$. A temperature fluctuation, $\Delta T$, vertically displaces the GVD curve according to the material thermo-optic coefficient. Simultaneously, the correlated change in $\lambda_{\rm{CS}}$ causes $f_{\rm{rep}}$ to move along its new GVD curve. Hence, $\eta_T$ is calculated from the total vertical displacement, $\Delta f_{\rm{rep}}$, divided by $\Delta T$, for small $\Delta T$. In the next section, we apply simulation techniques to make these concepts more precise. 

\begin{figure}[b]
    \centering
    \includegraphics[width=\linewidth]{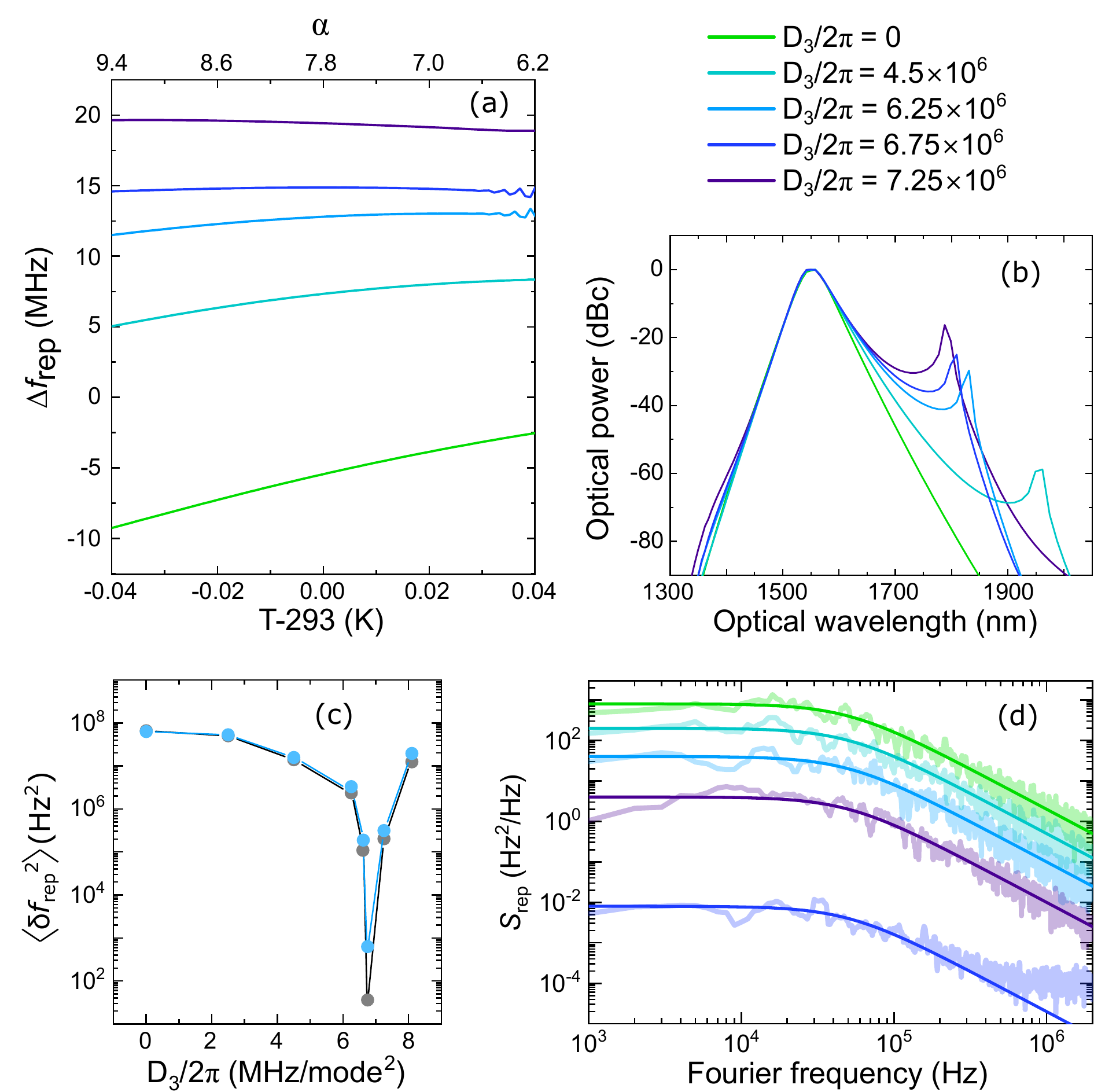}
    \caption{LLE simulations of $\eta_T$ and soliton-microcomb thermal noise for various GVD settings. Parameters for the simulations are: $D_{\rm{2}}/2\pi=60$ MHz/mode, $F^2=10$, and $\alpha=7.65$, (a) Simulated $f_{\rm{rep}}$ versus $T$. As $D_{\rm{3}}$ is increased from zero, $\eta_T$ decreases and eventually becomes negative. (b) Optical spectra for various $D_{\rm{3}}$. (c) Variance of $f_{\rm{rep}}$ frequency fluctuations, $\left< \delta f_{\rm{rep}}^2 \right>$, calculated in two ways: From the slopes in (a) (gray data points) and by integrating $S_{\rm{rep}}$ (blue data points). (d) Simulated $S_{\rm{rep}}$ spectra for various $D_{\rm{3}}$. Faded lines are LLE simulation results, and bold lines are fits to the data.} 
    \label{fig:D3}
\end{figure}

\begin{figure*}
    \centering
    \includegraphics[width=\linewidth]{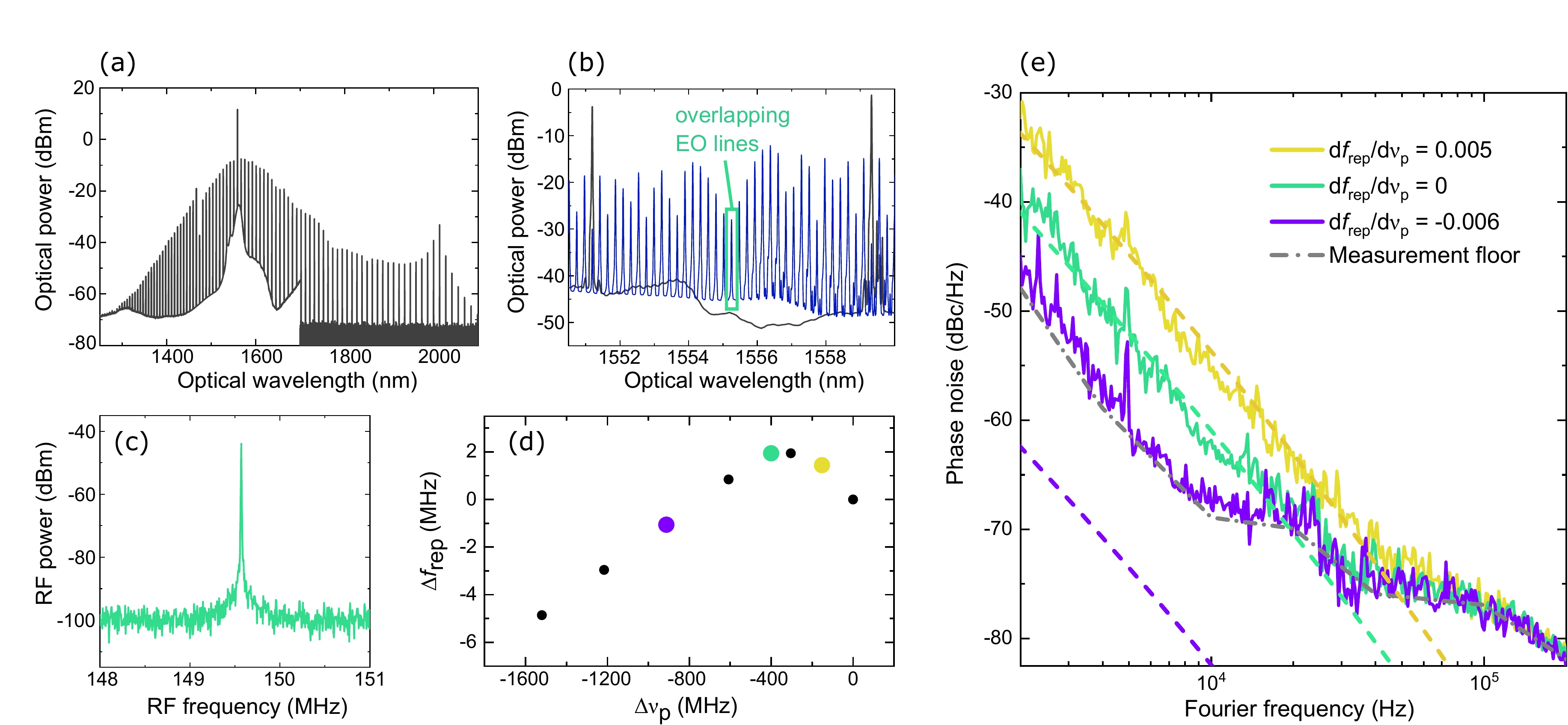}
    \caption{Experimental evidence for thermal-noise mitigation by a balance of thermal shifts in GVD and $\Omega$. (a) Optical spectrum used in the experiments. (b) Repetition frequency ($f_\text{rep}$) detection by electro-optic modulation. (c) RF spectrum of the thermal-noise-limited $f_{\rm{rep}}$ signal. (d) $f_{\rm{rep}}$ versus $\nu_{\rm{p}}$. These data are used to estimate $\eta_T$ and predict $S_{\rm{rep}}$. (e) $S_{\rm{rep}}$ phase-noise measurements of the 1-THz repetition rate for various settings of $\nu_{\rm{p}}$. Dashed lines correspond to predictions from our thermal noise model using $\eta_T$ estimates from (d).}
    \label{fig:experiment}
\end{figure*}

We model the temperature-dependent intraresonator field, $\psi$, using the Lugiato-Lefever equation (LLE) \cite{coen2013modeling, chembo2013spatiotemporal}, including self-steepening (SS) and an approximation \cite{karpov2016raman} for stimulated Raman scattering (SRS), 
\begin{equation}{\label{eq:LLE}}
\frac{\partial \psi}{\partial t}=F-(1+i\alpha) \psi-i\mathcal{D}\tilde{\psi}+(i+(\theta_{k}-i\theta_{R})\frac{\partial}{\partial\theta})\lvert \psi \rvert^2 \psi,
\end{equation}
where $F^2$ is the normalized pump-laser power, $\alpha(T)=\frac{2}{\Gamma} \times (\nu_{\rm{0}}(T)-\nu_{\rm{p}})$ is the temperature-dependent detuning between the pump-laser frequency, $\nu_p$, and the microresonator mode frequency, $\nu_{\rm{0}}(T)$, that is normalized to the modal linewidth, $\Gamma$; $\mathcal{D}(\mu, T)=\frac{2}{\Gamma}\times (\nu_{\mu}(T)-\nu_{\rm{0}}(T)-\frac{\mu D_{\rm{1}}}{2\pi})$ is the temperature-dependent microresonator dispersion, where $\mu$ is the mode number with respect to $\nu_{\rm{0}}$ and $\frac{D_{\rm{1}}}{2\pi}$ is the microresonator free-spectral range (FSR); $\tilde{\psi}$ represents that operations to the intraresonator field are performed in the frequency domain; $\theta_{k}$ and $\theta_{R}$ are coefficients related to SS and SRS, respectively \cite{bao2017soliton,bao2015soliton,karpov2016raman}; and $\theta=\tau D_{\rm{1}}$ is a fast-time variable corresponding to the intraresonator angle in a moving reference frame. All of the soliton microcombs in our study have a 3-dB spectral bandwidth that is $<10$ THz, so that only the instantaneous Raman response is included in the model \cite{karpov2016raman}. Furthermore, we distinguish the temperature-dependent dispersion, $\mathcal{D}$, from the integrated dispersion, $D_{\rm{int}}=2\pi\times (\nu_{\mu}-\nu_{\rm{0}}-\frac{\mu D_{\rm{1}}}{2\pi})=\sum_{j\geq2}D_{j}\mu^j/j!$, and calculate $\nu_{\mu}(T)$ to first order as 
\begin{equation}{\label{eq:tempco}}
\begin{split}
\nu_{\mu}(T)&=\nu_{\mu}(T_{\rm{0}})+(T-T_{\rm{0}})\frac{d\nu_{\mu}}{dT}\\
&=\nu_{\mu}(T_{\rm{0}})-(T-T_{\rm{0}})\times \eta_{\nu}\frac{\nu_{\mu}^2(T_{\rm{0}})L_{RT}}{(\mu+m)c},
\end{split}
\end{equation}where $\eta_{\nu}=\frac{dn_{\rm{p}}}{dT}$ is the material thermo-optic coefficient for the refractive index, $n_{\rm{p}}$; $m$ is the mode number corresponding to $\nu_{\rm{0}}$; and $\nu_{\mu}(T_{\rm{0}})$ are defined by $D_{\rm{int}}$ and $\nu_{\rm{0}}(T_{\rm{0}})$. We calculate $f_{\rm{rep}}$ directly from our LLE simulations by monitoring the soliton position in the moving reference frame; however, an insightful approximation for $f_{\rm{rep}}$ is given by
\begin{equation}{\label{eq:frep}}
2\pi f_{\rm{rep}}= \frac{2\pi \,(\nu_{\rm{1}}-\nu_{\rm{-1}})}{2}+\Omega\frac{D_{\rm{2}}}{D_{\rm{1}}}, 
\end{equation}where $\Omega=2\pi \times (\nu_{\rm{CS}}-\nu_{\rm{p}})$ is the detuning-dependent shift of the soliton carrier-wave frequency, $\nu_{\rm{CS}}=c/\lambda_{\rm{CS}}$, that corresponds to asymmetry in the comb spectrum around $\nu_{\rm{p}}$. In general, asymmetries arise from GVD, SRS, and spectral recoil from dispersive waves or mode crossings \cite{yi2017single}. Hence, temperature shifts induce a response in the comb spectrum, $\Omega(T)$ or $\lambda_{\rm{CS}}(T)$, that is tunable through the microresonator GVD. To mitigate thermal noise, the GVD should be designed to optimize $\Omega(T)$ so that the two terms in Eq. \ref{eq:frep} have opposite temperature dependence.

In Figure \ref{fig:D3} we explore the fundamental connection between GVD and thermal noise. We simulate soliton microcombs for fixed $\nu_{\rm{p}}$, $F^2$, and $D_{\rm{2}}$, while varying $D_{\rm{3}}$. We choose to vary $D_{\rm{3}}$ because it is the lowest-order term in the $D_{\rm{int}}$ expansion that gives rise to spectral asymmetry. First, we sweep the temperature, $T$, from $293.05$ K to $292.95$ K and monitor $f_{\rm{rep}}$, as shown in Fig. \ref{fig:D3}a. With $D_{\rm{3}}=0$, the $f_{\rm{rep}}$ tuning is dominated by SRS, which is known to exhibit $\frac{d f_{\rm{rep}}}{d \alpha}<0$ \cite{yi2016theory}; here this manifests as large $\eta_T > 0$. As $D_{\rm{3}}$ is increased from zero, $\eta_T$ decreases and eventually becomes negative. From these data and for specific values of $T$ we can calculate $\eta_T$ and compare the value $\lvert \eta_T \rvert^2 \frac{k_B T^2}{\rho C V}$ to the simulated noise variance, $\left< \delta f_{\rm{rep}}^2 \right>$; see Fig. \ref{fig:D3}c. Discrepancies between these two calculations indicate the importance of higher-order corrections to $\eta_T$ (i.e. it quantifies contributions to $\eta_T$ stemming from the curvature of the data in Fig. \ref{fig:D3}a). To calculate $\left< \delta f_{\rm{rep}}^2 \right>$ and gain a more comprehensive picture of thermal noise, we simulate the noise power spectral density of $f_{\rm{rep}}$ frequency fluctuations, $S_{\rm{rep}}$, by including temperature within our model as a stochastic variable, subject to fluctuation dissipation
\begin{equation}{\label{fluctuation_dissipation}}
\dot{T}=-\Gamma_T \Delta T+\zeta_T,
\end{equation}where $\Gamma_T$ is the thermal dissipation rate and $\zeta_T$ is a fluctuation source defined by its autocorrelation, $\left< \zeta_T(t) \zeta_T(t+\tau) \right>=\frac{2\Gamma_T k_B T^2}{\rho C V}\rm{\delta}(\tau)$, where $\rm{\delta}(\tau)$ is the Dirac $\rm{\delta}$ function \cite{sun2017squeezing}. Remarkably, for the same magnitude of thermal noise present in the microresonator, we observe a $>50$ dB suppression of $f_{\rm{rep}}$ frequency fluctuations, as shown in Fig. \ref{fig:D3}d. Such unprecedented flexibility in the thermal noise limit is a direct result of the nonlinear, multi-mode nature of soliton microcombs. 

We perform experiments to test our modeling and predictions, using a single soliton circulating a $\rm{Si}_3\rm{N}_4$ (SiN) microresonator at a rate $f_{\rm{rep}}\approx1 $ THz; the optical spectrum is pictured in Fig. \ref{fig:experiment}a. We measure $f_{\rm{rep}}$ by electro-optic modulation \cite{drake2018kerr}, as shown in Fig. \ref{fig:experiment}b. We confirm a thermal-noise-limited $f_{\rm{rep}}$ signal by ruling out fluctuations of both the pump-laser frequency and intensity \cite{drake2019thermal} and by comparing our measurements to a thermal noise model \cite{kondratiev2018thermorefractive}. In our experiments, we cannot accurately control the modal temperature; therefore, we record $f_{\rm{rep}}$ versus $\nu_{\rm{p}}$ (Fig. \ref{fig:experiment}d) and understand $\eta_T$ through the decomposition
\begin{equation}{\label{eq:etaT}}
\eta_T=\frac{df_{\rm{rep}}}{dT}=\frac{d\nu_{\rm{0}}}{dT}\left(\frac{\partial f_{\rm{rep}}}{\partial \nu_{\rm{0}}}+\frac{2}{\Gamma}\frac{\partial f_{\rm{rep}}}{\partial \alpha}\right),
\end{equation}
and approximate that for our measurements, $\frac{\Gamma}{2}\frac{d\alpha}{d\nu_{\rm{p}}} \approx -1$, which is valid for large $\alpha$ \cite{stone2018thermal}. Moreover, for SiN microresonators, $\frac{d\nu_{\rm{0}}}{dT} \approx -2.5$ GHz/K \cite{Huang2019TRnoise,drake2019thermal} and $\frac{\partial f_{\rm{rep}}}{\partial \nu_{\rm{0}}} \approx 1/m$ \cite{xue2016thermal}. Hence, we estimate $\eta_T$ from our measurements as $-2.5 \frac{\rm{GHz}}{\rm{K}} (5.2 \frac{\rm{MHz}}{\rm{GHz}} - \frac{df_{\rm{rep}}}{d\nu_{\rm{p}}})$. Importantly, Eq. \ref{eq:etaT} predicts that for a thermal-noise-limited $f_{\rm{rep}}$ signal, operating near $\frac{df_{\rm{rep}}}{d\nu_{\rm{p}}} = 0$ does not yield the lowest noise as for previous observations of so-called "quiet points" \cite{yi2017single}. Rather, the thermal noise will be mitigated significantly when $\frac{df_{\rm{rep}}}{d\alpha} \approx -\frac{df_{\rm{rep}}}{d\nu_{\rm{0}}}$, which physically corresponds to a balance between thermal changes in the FSR with thermal changes in $\Omega$ (i.e. a balance in the two terms of Eq. \ref{eq:frep}). Guided by Eq. \ref{eq:etaT}, we estimate $\eta_T$ values for the purple, green, and gold data points as $0.8$, $-13$, and $-30$ MHz/K, respectively. Our phase-noise measurements are consistent with these values and show $\approx 15$ dB of noise suppression for the different settings of $\nu_{\rm{p}}$, but the lowest-noise data is limited by our measurement floor, which is set by the synthesizer used to drive the electro-optic modulators for $f_{\rm{rep}}$ detection. The two phase-noise traces above the measurement floor agree with our thermal-noise model, shown by the dashed lines in Fig. \ref{fig:experiment}e. Based on our estimations of $\eta_T$, we expect that $f_{\rm{rep}}$ frequency fluctuations are suppressed by almost $30$ dB when operating at $\Delta \nu_{\rm{p}}=900$ MHz (purple data) compared to $\Delta \nu_{\rm{p}}=180$ MHz (gold data). Our measurements confirm that thermal noise in soliton microcombs is not a rigid limit set by material properties, but instead arises from complex interactions between many microcomb modes as determined by optical nonlinearity (especially SRS) and GVD. 
\begin{figure}
    \centering
    \includegraphics[width=\linewidth]{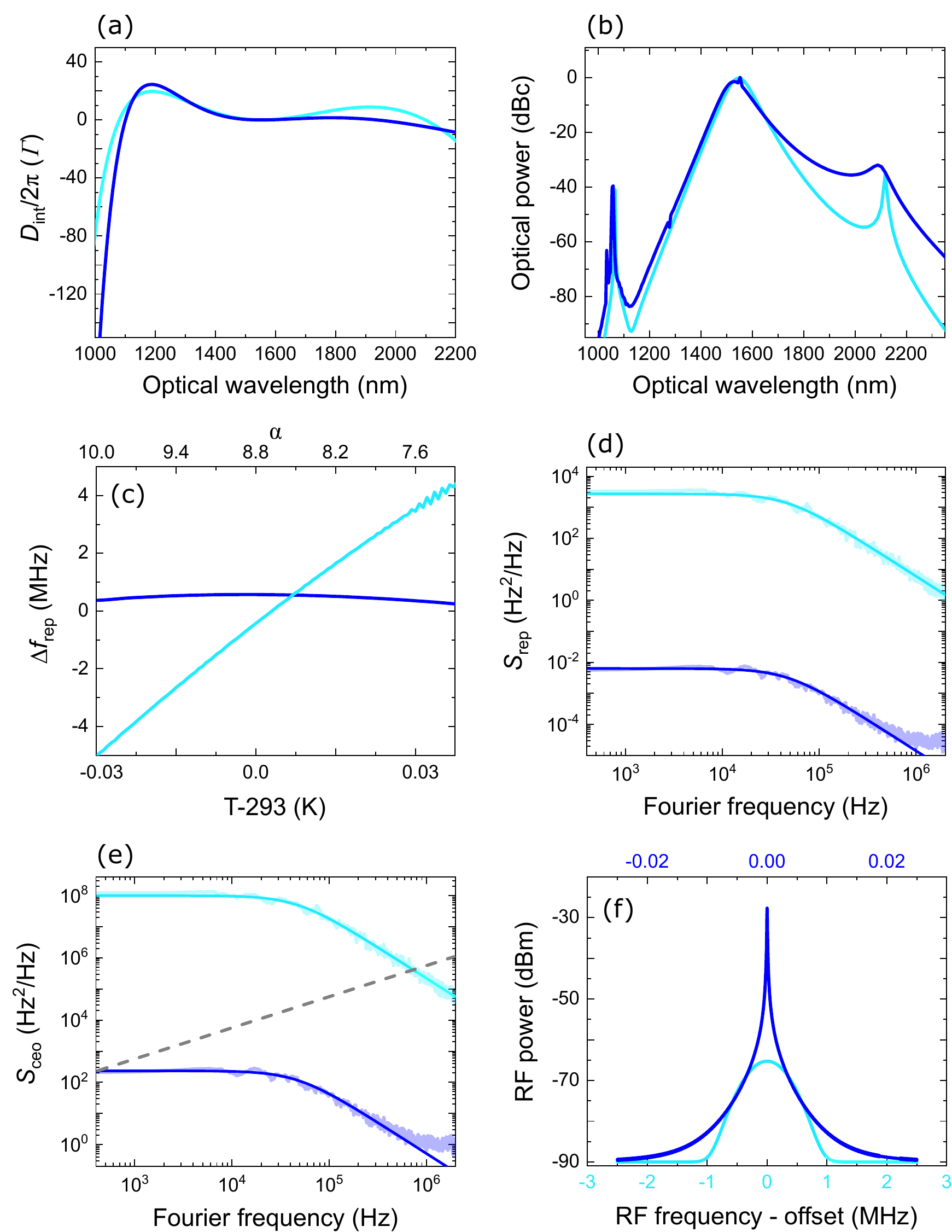}
    \caption{Simulations of thermal noise in octave-spanning soliton microcombs. (a) $D_{\rm{int}}$ for the two solitons analyzed in this study. Parameters for the dark blue traces (units omitted): $D_{\rm{2}}/2\pi=20\times 10^6$, $D_{\rm{3}}/2\pi=1.5\times 10^6$, $D_{\rm{4}}/2\pi=-52\times 10^3$, $D_{\rm{5}}/2\pi=-4\times 10^3$, $F^2=12$, $\alpha=9.3$. Parameters for the light blue traces: $D_{\rm{2}}/2\pi=20\times 10^6$, $D_{\rm{3}}/2\pi=0$, $D_{\rm{4}}/2\pi=-120\times 10^3$, $D_{\rm{5}}/2\pi=5.3\times 10^3$,  $D_{\rm{6}}/2\pi=-150$, $F^2=12$, $\alpha =9.25$. (b) Optical spectra corresponding to the dispersion profiles in (a). (c) $f_{\rm{rep}}$ versus $T$. (d) $S_{\rm{rep}}$ spectra. (e) $S_{\rm{ceo}}$ spectra calculated from $m^2 \times S_{\rm{rep}}$. The dashed line is the so-called beta line for understanding which Fourier frequencies contribute to the signal linewidth. (f) Simulated $f_{\rm{ceo}}$ beatnote that indicates both coherence and signal-to-noise ratio are improved by mitigating thermal noise. The top (bottom) axis refers to the blue (cyan) trace. } 
    \label{fig:OctaveCombs}
\end{figure}

Finally, we model the impact of thermal noise on octave-spanning soliton microcombs and emphasize its role in $f_{\rm{ceo}}$ detection. First, we assess that $D_{\rm{3}}$ plays the primary role in coupling $D_{\rm{int}}$ to $\eta_T$; therefore, we model two spectrally similar, octave-spanning solitons with $D_{\rm{3}}/2\pi$ values of $0$ and $1.5\times10^6$ MHz/mode$^2$, respectively. $D_{\rm{int}}$ curves and optical spectra for each comb are shown in Figs. \ref{fig:OctaveCombs}a and \ref{fig:OctaveCombs}b, respectively. To approximately match the DW locations, we manipulate higher-order dispersion terms in the LLE. Importantly, for spectrally-broad solitons featuring strong DWs, DW recoil and soliton self-interactions can significantly impact $\eta_T$ \cite{skryabin2017self}. In our simulations, we have tried to avoid this regime by operating at low $F^2$, but note that understanding these effects will be important for future experiments. In Fig. \ref{fig:OctaveCombs}, we present simulation results comparing the two soliton microcombs. Despite having similar optical spectra, $\eta_T\approx 130$ MHz/K for the $D_{\rm{3}}=0$ comb (hereafter referred to as the noisy comb), indicating that SRS primarily controls the $f_{\rm{rep}}$ tuning, while $\eta_T \approx 0$ for the $D_{\rm{3}}/2\pi=1.5\times10^6$ comb (hereafter referred to as the quiet comb), as shown in Fig. \ref{fig:OctaveCombs}c. These dynamics are in agreement with the simpler combs analyzed in Fig. \ref{fig:D3}. Unsurprisingly, we observe a $>50$ dB difference in the $S_{\rm{rep}}$ spectra of the two combs. To understand the implications for $f_{\rm{ceo}}$, we calculate the noise power spectral density of $f_{\rm{ceo}}$ frequency fluctuations, $S_{\rm{ceo}}$, as $S_{\rm{ceo}}=m^2\times S_{\rm{rep}}$; the resulting spectra are shown in Fig. \ref{fig:OctaveCombs}e. These data have important implications for soliton-microcomb applications. For example, by comparing each spectrum with the beta line \cite{di2010simple}, we assess that stabilization of the soliton microcombs for coherent applications \cite{carlson2018ultrafast} would require $\approx 700$ kHz of servo bandwidth for the noisy comb (in addition to significantly greater gain for overcoming the excess noise) but only $\approx 400$ Hz of bandwidth for the quiet comb. Moreover, we integrate the spectra from Fig. \ref{fig:OctaveCombs}e and apply the Wiener-Khintchine Theorem to analyze the $f_{\rm{ceo}}$ beatnote. We make two noteworthy observations: First, the thermal-noise-limited linewidths for the noisy and quiet combs are approximately $1$ MHz and $100$ Hz, respectively. Second, the signal-to-noise ratios for the noisy and quiet combs are approximately $25$ dB and $60$ dB, respectively. Both of these measures impact applications requiring low noise and good optical coherence. 

In conclusion, soliton microcombs offer a unique lens through which to view thermodynamic processes and noise. We have shown how interactions between comb modes induce thermodynamic correlations that may be harnessed to manipulate the thermal noise limit. Our results shed light on the relationship between nonlinear physics in multi-mode systems and the intrinsic, microscopic fluctuations therein.  

\par We thank Jizhao Zang and Su Peng Yu for a careful reading of the paper. This project is funded by the DARPA DRINQS program. This work is a contribution of the U.S. government and is not subject to copyright in the US.

\bibliography{Bibliography}

\end{document}